\documentclass[12pt]{article}

\begin{document}
\title {{\bf 2D(1+1) Quantum Gravity. Gravitational
Quantum Stationary Hamilton-Jacobi Equation}}
\author { T. Djama }
\date {June 27, 2004}

\maketitle

\begin{abstract}
\noindent In the present article, we construct a 2D formulation of
quantum gravity in the framework of a deterministic theory. In
this context, a Quantum stationary Hamilton-Jacobi equation is
derived from the Klein-Gordon equation written in the presence of
a gravitational field. We show that this equation reduces to the
Quantum stationary Hamilton-Jacobi equation when the
gravitational field is not present in the 2D time-space. As a
second step, we introduce the quantum gravitational Lagrangian for
the quantum motion of a particle moving in the presence of a
gravitational field. We, deduce the relationship between the
gravitational quantum conjugate momentum and the velocity of the
particle.
\end{abstract}

\vskip\baselineskip



\newpage

\vskip0.5\baselineskip \noindent { {\bf 1.\ Introduction} }
\vskip0.5\baselineskip

Since three years, a deterministic approach of quantum mechanics
is presented in Refs.
\cite{BD1,BD2,BD3,Djama1,Djama2,Djama3,Djama4,Djama5} as an
alternative of the standard quantum mechanics (Copenhagen
interpretation). It consists on the introduction of a Quantum
Stationary Hamilton-Jacobi Equation (QSHJE) already established
by Faraggi and Matone \cite{FM1,FM2}, Floyd \cite {Flo1}, Bohm
\cite {Bohm1} and de Broglie \cite {Brog1}
$$
{1 \over 2m_0}\left({\vec{\nabla}S_0(\vec r)}\right)^2-{\hbar^2
\over 2m_0}{\Delta R(\vec r) \over R(\vec r)}+(E-V(\vec r))=0\;
\; , \hskip15mm   (a)
$$
\begin{equation}
\vec{\nabla} . \left(R^2(\vec r)\vec{\nabla}S_0\right)=0 \; \; ,
\hskip65mm (b)
\end{equation}
which correspond in classical mechanics $(\hbar \to 0)$ to the Classical
Stationary Hamilton-Jacobi Equation (CSHJE)
\begin{equation}
{1 \over 2m_0}\left({\vec{\nabla}S_0(\vec r)}\right)^2+(E-V(\vec
r))=0\; \; .
\end{equation}
$R$ and $S_0$ being real functions and $S_0$ representing the
action of the quantum system.

Before studying the 3D systems, we have investigated the 1D
problems, so, we took up the 1d QSHJE written as \cite {BD1,Flo1,Flo2}
\begin{eqnarray}
{1 \over 2m_0}\left({dS_0 \over dx}\right)^2- {\hbar^2 \over
4m_0}\left[{3 \over 2} \left({dS_0 \over dx}\right)^{-2}
\left({d^2 S_0 \over dx^2}\right)^2- \right. \hskip25mm\nonumber\\
\left. \left({dS_0 \over dx}\right)^{-1} \left({d^3 S_0 \over
dx^3}\right) \right]+ V(x)=E\; .
\end{eqnarray}
Taking advantage of the solution of the QSHJE (Eq. (3)) \cite
{BD1,Flo1,Flo2}
\begin{equation}
S_0(x)=\hbar \arctan\left(a\ {\theta(x) \over \phi(x)}+b\right)
\; ,
\end{equation}
we introduced a quantum Lagrangian of the form \cite {BD1}
\begin{equation}
L(x,\dot{x},a,b)={1\over 2} m {\dot{x}}^2 f(x,a,b)-V(x) \; ,
\end{equation}
and derived the dynamical relation giving the velocity of the
particle in function of the energy, the potential and the quantum
conjugate momentum \cite {BD1}
\begin{equation}
 \dot{x}\, {\partial S_0 \over \partial x}= {2(E-V)} \; ,
\end{equation}
Using this relation, we derived the first integral of the Quantum
Newton's Law \cite{BD1}, and plotted quantum trajectories of
particles moving under different potentials \cite{BD2}. Later, we
presented a generalization of our approach to the relativistic
systems \cite{Djama1,Djama2} and spinning particles
\cite{Djama3}. For the quantum relativistic systems we introduced
the Relativistic QSHJE (RQSHJE) \cite{Djama1}
\begin{eqnarray}
{1 \over 2m_0}\left({\partial S_0 \over \partial x}\right)^2-
{\hbar^2 \over 4m_0}\left[{3 \over 2} \left({\partial S_0 \over
\partial x}\right)^{-2} \left({\partial^2 S_0 \over \partial
x^2}\right)^2-\right.
\hskip35mm&& \nonumber\\
\left.\left({\partial S_0 \over \partial x}\right)^{-1}
\left({\partial^3 S_0 \over \partial x^3}\right) \right]+ {1
\over 2m_0c^2}\left[m_0^2c^4 -(E-V)^2\right]=0\; ,
\end{eqnarray}
and construct the Lagrangian of height energy particles as follows \cite{Djama1}
\begin{equation}
L(x, \dot{x},a,b)=-m_0c^2 \sqrt{1-f(x,a,b)\, {\dot{x}^2 \over c^2}}-V(x)\;
.
\end{equation}
The solution of Eq. (7) can be expressed by Eq. (4), where
$\theta$ and $\phi$ represent now two real independent solutions
of the Klein-Gordon equation
\begin{equation}
-c^2 \hbar^2 {\partial^2 \phi \over \partial x^2}+
\left[m_0^2c^4-(E-V)^2\right]\phi(x)=0\; .
\end{equation}
Applying the least action principle to the Lagrangian (8) and
after some calculations we leads to the dynamical relation
\cite{Djama1}
\begin{equation}
\dot{x}\; {\partial S_0 \over \partial x}={E-V(x)}- {m_0^2 c^4
\over (E-V)} ,
\end{equation}
from which we derived the relativistic quantum Newton's Law \cite{Djama1}. Then, in
Ref. \cite{Djama2}, we plotted the relativistic quantum trajectories for
different potentials.

The spinning particles are also studied in the context of our
deterministic approach of quantum mechanics. So, beginning with
the 1D Dirac's equation \cite{Djama3}
\begin{equation}
-i\; \hbar c \ \ \sigma_{x} \ \ {d \psi \over dx} \ =
(E-V(x)-\sigma_{z}\ m_{0}c^2) \ \psi
\end{equation}
where
\begin{equation}
\sigma_{x}=\begin{pmatrix}
      { 0&1  \cr
        1&0  \cr}
 \end{pmatrix}; \ \
 \sigma_{y}=\begin{pmatrix}
      { 0&-i  \cr
        i&\ 0  \cr}
 \end{pmatrix}; \ \
 \sigma_{z}=\begin{pmatrix}
      { 1&\ 0  \cr
        0&-1  \cr}
 \end{pmatrix}\, ,
\end{equation}
we established two RQSHJES$_{{1 \over 2}}$, which can be written
as \cite{Djama3}
\begin{eqnarray}
{1 \over 2m_0}\left({dS_{m_s} \over dx }\right)^2- {\hbar^2 \over
4m_0} \{S_{m_s},x\}+ {\hbar^2 \over 2m_{0}}(E-V+(-1)^{2\, m_s+1}\,
m_{0}c^2)^{1 \over 2}\ .
\hskip5mm&& \nonumber\\
\ .{d^2 \over dx^2} \left[(E-V+(-1)^{2\, m_s+1}\, m_{0}c^2)^{-{1
\over 2}}\right] + {1 \over 2m_0c^2}\left[m_0^2c^4
-(E-V)^2\right]=0\; ,
\end{eqnarray}
where $S_{m_s}$ represents the quantum reduced action of the
particle with spin ${1 \over 2}$ corresponding to the projection
$m_s=\pm {1 \over 2}$ of the spin \cite{Djama3}. The quantity
$$
\{S_{m_s},x\}=\left[{3 \over 2} \left({dS_{m_s} \over dx
}\right)^{-2} \left({d^2 S_{m_s} \over
dx^2}\right)^2-\left({dS_{m_s} \over dx}\right)^{-1} \left({d^3
S_{m_s} \over dx^3}\right) \right]
$$
represents the Schwarzian derivative of $S_{m_s}$ with respect to
$x$.

All the above results are presented for the 1D spaces. However,
the physical phenomena happens in the 3D spaces. This is the
reason why we present in Refs. \cite{Djama4,Djama5} a generalization to 3D of our
deterministic approach. First, we have presented a solution of
the 3D QSHJE (Eqs. (1)) and given the reduced action by the
relation (4), but where $\theta$ and $\phi$ are two real
independent solutions of the 3D Schr\"odinger equation. Secondly,
we introduced a 3D quantum Lagrangian \cite{Djama4}
\begin{equation}
L_q={m_0 \over 2}\dot {x}^2a_{xx}(\vec{r})+{m_0 \over
2}\dot{y}^2a_{yy}(\vec{r})+{m_0 \over 2}\dot{z}^2a_{zz}(\vec{r})-
V((\vec{r}))\; ,
\end{equation}
from which we derived the dynamical relation connecting between
the quantum conjugate momentum and the velocity of the particle \cite{Djama3}
\begin{equation}
\vec{v}\; . \; \vec{\nabla S_0}=2\left[E-V(\vec{r})\right]\; .
\end{equation}
In Eq. (14), the quantities $a_{\mu\mu}$ represents the diagonal
components of the quantum metric tensor which define the
curvature of the space by the quantum potential.
The components $a_{\mu\mu}$ play the same role as the function
$f(x)$ present in Eq. (5) for the 1D systems. Using Eq. (15) after
taking advantage on the expression of the reduced action $S_0(\vec{r})$
(Eq. (4)) and the solutions of the 3D
Schr\"odinger equation, we plot the quantum trajectories of the
Hydrogen's electron for different states \cite{Djama4}.

Remark that in our formulation, we obtain easily the classical
equations only by taking $\hbar \to 0$ in the quantum equations
(1-15). This is an important result, since we build a
deterministic quantum mechanics which, at the classical limit
($\hbar \to 0$), leads to the deterministic classical mechanics
\cite{BD1,BD2,BD3,Djama1,Djama2,Djama4,Djama5}.

After generalizing our approach to the relativistic systems, the
spinning particles and the 3D spaces, an interesting questions
arise. It concerns the gravity interaction. Is it possible to
study the motion of a quantum particle in the presence of a
gravitational field? If yes, how it should be made? The answer of
these questions may construct a deterministic approach of quantum
gravity.

Since the appearance of quantum mechanics many theories of
quantum gravity are developed in the context of the standard
quantum mechanics ie the probabilistic approach of quantum
phenomena. String theory, Dilaton theories and String-dilaton
theories are the most important examples of such theories
\cite{Mann1,Mann2,Mann3,Mann4,Mann5,Mann6,Mann7,Mandal,Witt,Napp,Call,Mann8,Teo,Mann9,Mann10}.
 In spite of their theoretical results, many difficulties stand
in the way of these theories. One of these difficulties  consist
on the fact that these theories connect between a probabilistic
quantum theory and a deterministic theory of gravity (General
Relativity). This difficulty is a conceptual one and not
mathematical. In fact, the gravitational field which is a
manifestation of a mass $M$ present in a determined space
(Newton's theory of gravity, General Relativity) is now described
by a quantum theory which stipulate that the mass $M$ do not
occupy a well determined space but is spread over all the space
according to a probabilistic law. However, when we take the
classical limit, one should find the deterministic theory of
gravitation (GR), so, we don't know how a probabilistic theory
should reduce to a deterministic one at the classical limit. Thus,
in our opinion, it is necessary to build a deterministic quantum
gravity theory connecting between two deterministic theories, GR
and a deterministic quantum mechanics. For this aim, we propose in
this article to connect our deterministic approach of quantum
mechanics and the general relativity.

Now, let us remind the most results of general relativity
\cite{EIN,Dirac}. After announcing the equivalence postulate,
Einstein deduce that in order to generalize the Poisson equation
of the gravitational field $\Phi$
\begin{equation}
\Delta \Phi-4 \pi G \mu=0
\end{equation}
to the relativistic theory, the space must be curved. After
setting the interval $ds$ as
\begin{eqnarray}
ds^2=g_{\mu\nu}\; dx^{\mu}dx^{\nu}\hskip15mm \mu, \, \nu=0,\, 1,\,
2,\, 3\; ,
\end{eqnarray}
Einstein obtained the following field equations \cite{EIN,Dirac}
\begin{eqnarray}
E_{\mu\nu}=R_{\mu\nu}-g_{\mu\nu}R={8 \pi G \over
c^4}T_{\mu\nu}\hskip15mm \mu, \, \nu=0,\, 1,\, 2,\, 3\;
\end{eqnarray}
$G$ being the universal gravitational constant, $\mu$ is the mass
density and $g_{\mu\nu}$ is the metric tensor of the curved
space. $T_{\mu\nu}$ is the momentum-energy tensor. $R_{\mu\nu}$
is the Ricci tensor given as \cite{EIN,Dirac}

\begin{equation}
R_{\mu\nu}=\partial_\gamma\,
\Gamma^{\gamma}_{\mu\nu}-\partial_\nu\,
\Gamma^{\gamma}_{\mu\gamma}
+\Gamma^{\gamma}_{\mu\nu}\Gamma^{\delta}_{\gamma\delta}-
\Gamma^{\delta}_{\mu\gamma}\Gamma^{\gamma}_{\nu\delta}\; .
\end{equation}
$\Gamma^{\gamma}_{\mu\nu}$ are the Christoffel symbols
\begin{equation}
\Gamma^{\gamma}_{\mu\nu}={1 \over
2}g^{\gamma\delta}\left({\partial g_{\delta\mu} \over
\partial x^\nu}+{\partial g_{\delta\nu} \over
\partial x^\mu}-{\partial g_{\mu\nu} \over
\partial x^\delta}\right)\; .
\end{equation}
$R$ is the curvature invariant
\begin{equation}
R=g^{\mu\nu}R_{\mu\nu}\, ,\hskip15mm \mu, \, \nu=0,\, 1,\, 2,\,
3\; .
\end{equation}
The free particle describe a trajectory given by the geodesic
equations \cite{EIN,Dirac}
\begin{equation}
{d^2 x^{\sigma} \over ds^2}+\Gamma_{\mu\nu}^\sigma {d x^{\mu}
\over ds}{d x^{\nu} \over ds}=0\; , \hskip17mm \mu, \, \nu=0,\,
1,\, 2,\, 3\; .
\end{equation}
In fact, the last equation represents the law of motion under the
gravitational field and can be derived from the Einstein-Hilbert
action \cite{EIN,Dirac}
\begin{equation}
S(x^{\mu})=-m_0c\int ds\, = -m_0c\int \sqrt{g_{\mu\nu}\dot{x}^{\mu}\dot{x}^{\nu}}\, ds
\end{equation}
after using the least action principle. $ds$ is the interval given in Eq. (17) and is
considered as an evolution parameter, and $\dot{x}^{\mu}=dx^{\mu}/ds$.

\vskip0.25\baselineskip The results of the GR will be taking into
consideration when we try to construct the quantum gravity
formulation. However, a construction of a quantum gravity theory
in 4D seems to be so complicated. Then, we propose to construct a
2D formulation of quantum gravity. In this order, we should first
investigate the 2D classical gravity, and determine the metric of
the 2D space. We will see in Sec. 2 the difficulties we face to
construct such a gravity theory. In Sec. 3, in spite of the
difficulties we face in Sec. 2, we derive a Gravitational QSHJE
(GQSHJE) from the 2D Klein-Gordon equation written in the
presence of the gravitational field. After, in Sec. 4, we
introduce a Lagrangian from which we derive the equation of
motion of a free particle for a stationary system and a static
metric . Finally, in Sec. 5, we introduce the Einstein-Hilbert
action and derive the quantum gravitational law of motion for a
non-stationary system and a general metric.

\vskip0.5\baselineskip \noindent { {\bf 2.\ General relativity
and 2D Gravity } } \vskip0.5\baselineskip

In 2-dimensions, the Einstein's gravitational theory is trivial.
This is due to the fact that the Einstein tensor
$E_{\alpha\beta}$ (Eq. (18)) is identically zero for all
2-dimensional metrics. As a consequence, the Einstein equations
(Eq. (18)) indicate that the energy-momentum tensor is vanished
and this result is inconsistent with some non-trivial matter
configurations. This problem is the main obstacle which the
physicists face with. Then, what is the constraint which will
define the metric of the 2 dimensional space-time in the presence
of the gravitational field? As an answer, many interesting
suggestions are advanced
\cite{Mann1,Mann2,Mann3,Mann4,Mann5,Mann6,Mann7,Mandal,Witt,Napp,Call,Mann8,Teo,Mann9,Mann10}.
One of these suggestions has shown that classical gravity in two
spacetime dimensions need not be so trivial \cite {Mann1}, and
that an interesting relativistic theory of gravitation in this
context may be formed by setting
\begin{equation}
 R =8\pi \, GT=8\pi \, GT_{\mu}^{\mu}\, ,
\end{equation}
$T=T_{\mu}^{\mu}$ representing the trace of the conserved energy-momentum tensor.
It is shown \cite {Mann10} that this theory is a 2D limit of the 4D General Relativity.
In spite of its simplicity, this approach has many interesting classical and semi-
classical results, including a well-defined Newtonian limit \cite {Mann1}, black holes
\cite {Mann2,Mann3}, a post-Newtonian expansion, gravitational waves, FRW cosmologies,
gravitational collapse \cite {Mann4} and black hole radiation
\cite {Mann4,Mann5,Mann6}.
The aim of the present paper is not to define the metric of the 2D spaces,
but is to construct a theory of a quantum gravity in the 2D spaces. Such a theory will
permit to have an idea about the manner with which we should construct the 4D theory
of quantum gravity.


\vskip1.\baselineskip \noindent { {\bf 3.\ Gravitational Quantum
Stationary Hamilton-Jacobi Equation} } \vskip0.25\baselineskip

\noindent {{\bf a.\ general form of the GQHJE } }
\vskip0.25\baselineskip
As we have mentioned above, in the present paper, we restrict our
study on the 2D spaces ie spaces with the temporal coordinate $t$
and one spatial coordinate $x$. In this section we derive the
(1+1)D Gravitational Quantum Hamilton-Jacobi Equation (GQHJE) from
the (1+1)D Klein-Gordon equation written in the case of the
presence of a gravitational field. The considered Klein-Gordon
equation is given in Ref. \cite {Dar} as
\begin{equation}
{1 \over \sqrt{|g|}}\, \partial_{\alpha}\, (\sqrt{|g|}\,
g^{\alpha\beta}\, \partial_{\beta})\, \Psi(X)+ { m^{2}_o\, c^2 \over \hbar^2}
\Psi(X)=0\, , \hskip8mm \alpha, \, \beta \, =0,\, 1\; .
\end{equation}
$\Psi(X)$ is the wave function, and $X:=(t,x)$. $g^{\alpha\beta}(X)$ are the components of
the inverse of the metric tensor $g(X)$. $|g|$ represents the determinant's positive value
of the metric tensor. Eq. (25) can be written as
\begin{equation}
g^{\alpha\beta}\, \partial^{2}_{\alpha\beta}\, \Psi\, +\,
\partial_{\alpha}\, (\sqrt{|g|}\,
g^{\alpha\beta})\, \partial_{\beta} \, \Psi+ { m^{2}_o\, c^2 \over \hbar^2}
\Psi=0\, ,
\, \, \,   \alpha, \, \beta \, =0,\, 1\; ,
\end{equation}
where $\partial^{2}_{\alpha\beta}$ is the second order derivative with respect
to $x^{\alpha}$ and $x^{\beta}$.

Now, we derive the GQHJE. Let us write the wave function $\Psi$
with the Bohm-de Broglie notation
\begin{equation}
\Psi(X)=A(X)\, . \exp\left({{i \over \hbar}\, S(X)}\right)\; ,
\end{equation}
where $A(X)$ and $S(X)$ are real functions representing the
amplitude and the phase of the wave function. Replacing Eq. (27)
into Eq. (26) and separating the imaginary and real parts of the
resulting equation, we get
\begin{eqnarray}
g^{\alpha\beta}\, \partial_{\alpha}S\, \partial_{\beta}S
-{\; \hbar^2 \over A}\, g^{\alpha\beta}\, \partial^{2}_{\alpha\beta}\, A\,
-\, {\; \; \hbar^2 \over A\, \sqrt{|g|}}\, .
\hskip20mm\nonumber \\
.\, \partial_{\alpha}\, (\sqrt{|g|}\, g^{\alpha\beta})\,
\partial_{\beta}\, A\, -\, m^{2}_o\, c^2
=0
\end{eqnarray}
and
\begin{equation}
g^{\alpha\beta}\, (\partial_{\beta}A\partial_{\alpha}S\, +\,
\partial_{\alpha}A\partial_{\beta}S+A\, \partial^{2}_{\alpha\beta}\, S)+
{1 \over \sqrt{|g|}}\, \partial_{\alpha}\, (\sqrt{|g|}\, g^{\alpha\beta})\,
A\, \partial_{\beta}\, S
=0
\end{equation}
The two last equations represent the GQHJE for a general metric
and non-stationary case. Note that when we take the limit $\hbar
\to 0$ Eq. (28) reduces to the Hamilton-Jacobi equation written
in the presence of a gravitational field \cite{EIN,Dirac}
\begin{equation}
g^{\alpha\beta}\, \partial_{\alpha}S\, \partial_{\beta}S
=\, m^{2}_o\, c^2\; .
\end{equation}
%
%
\vskip1.\baselineskip \noindent { {\bf b.\ The GQSHJE for a
static metric and stationary case} } \vskip0.25\baselineskip
Now, we investigate the static metric and the stationary case.
The general form of a static metric is given by ($g_{1o}=g_{o1}=0$)
\begin{eqnarray}
g=\begin{pmatrix}
      { g_{oo}(x)&0  \cr
        0&g_{11}(x)  \cr}
 \end{pmatrix}; \hskip40mm\nonumber\\ \
g^{-1}=\begin{pmatrix}
      { g^{oo}(x)&\ \ 0  \cr
        0&g^{11}(x)  \cr}
 \end{pmatrix}=\begin{pmatrix}
      { {1\over g_{oo}(x)}&\ \ 0  \cr
        0&{1\over g_{11}(x)}  \cr}
 \end{pmatrix}, \ \
\end{eqnarray}
while for the stationary system we have
\begin{equation}
S(X)=S_0(x)-E\, t;\hskip15mm A(X)=A(x)\, .
\end{equation}
Replacing Eqs. (31) and (32) into Eqs. (28) and (29), we find after devising by $2m_o$
\begin{eqnarray}
{1 \over 2m_o}\left({\partial S_0\over \partial x}\right)^2\, {1 \over g_{11}}
-{\; \hbar^2 \over  2m_o\, g_{11}}\, {1 \over A}{\partial^2 A\over \partial x^2}-
{\; \; \hbar^2 \over 2m_o\, \sqrt{|g|}}\, .
\hskip33mm\nonumber \\
.\,
 {\partial \; \over \partial x}\, \left({\sqrt{|g|}\, \over g_{11}}\right)\,
 {1\over A}{\partial A\over \partial x}\,
 +{1 \over 2m_o\, c^2}\left({E^2 \over g_{oo}}-m^{2}_o\, c^4\right)
=0
\end{eqnarray}
and
\begin{equation}
{1 \over \sqrt{|g|}}{\partial\; \over \partial x}
\left[\sqrt{|g|}\, {1 \over g_{11}}\, A^2\, {\, \partial S_0 \over \partial x}\right]=0\, .
\end{equation}
Resolving EQ. (34), we find
\begin{equation}
A(x)=k\, \left({\sqrt{|g|}\, \over g_{11}}\,
{\, \partial S_0 \over \partial x}\right)^{-{1 \over 2}}
\end{equation}
with which after replacing into Eq. (33) we get
\begin{eqnarray}
{1 \over 2m_o}\left({\partial S_0\over \partial x}\right)^2\, {1 \over g_{11}}
-{\; \hbar^2 \over  4m_o\, g_{11}}\, \{S_0,x\}\,
+{\hbar^2 \over 2m_0\sqrt{g_{11}\, \sqrt{|g|}}}\, .
\hskip20mm\nonumber \\
.\, {\partial^2 \over \partial x^2}
\left(\sqrt{ {\sqrt{|g|}\over g_{11}}}\right)+
{1 \over 2m_o\, c^2}\left({E^2 \over g_{oo}}-m^2_o\, c^4\right)=0\; .
\end{eqnarray}
This last equation represents the GQSHJE for the static metric and the
stationary systems. We remark that after taking the limit $\hbar \to 0$
into Eq. (36), we find
\begin{equation}
{1 \over 2m_o}\left({\partial S_0\over \partial x}\right)^2\, {1 \over g_{11}}
+{1 \over 2m_o\, c^2}\left({E^2 \over g_{oo}}-m^2_o\, c^4\right)=0\; .
\end{equation}
It is clear that Eq. (37) is the classical Hamilton-Jacobi
equation in the presence of a gravitational field and represents
the particular case of Eq. (30) (Stationary system and static
metric).

Note also that if we consider a vanishing gravitational field
($g_{oo}=1$ and $g_{11}=-1$), Eq. (36) goes to the RQSHJE
(Eq. (7)).

\vskip1.\baselineskip \noindent { {\bf c.\ The solution of the
GQSHJE } } \vskip0.25\baselineskip

As a solution of the GQSHJE (Eq. (36)) we propose the expression (4) of the
reduced action, but where $\theta(x)$ and $\phi(x)$ represent
the two real and independent solutions of the Klein-Gordon
equation describing the case of a static metric and a stationary
systems
\begin{equation}
{\hbar^2c^2 \over g_{11}}\, {\partial^2 \psi(x) \over \partial x^2\; \; \; \, }+
{\hbar^2c^2 \over \sqrt{|g|}}\,
{\partial \psi(x) \over \partial x\; \; \; }\,
{\partial \over \partial x}\left({\sqrt{|g|} \over g_{11}}\right)\, +\,
\left(m_o^2c^4-{E^2 \over g_{oo}}\right)\, \psi(x)=0\, ,
\end{equation}
from which we deduce that the wronwkian of $\theta$ and $\phi$ verifies
\begin{equation}
W=\phi \, {d\theta\over dx}-\theta \, {d\phi\over dx}=\lambda\,
{g_{11}\over \sqrt{|g|}}\; .
\end{equation}
Now, we verify that expression (4) is the solution of Eq. (36).
In this order, let us write Eq. (4) in the following form
\begin{equation}
S_0(x)=\hbar \arctan\left({\theta^{'} \over \phi}\right)\; ,
\end{equation}
where
$$\theta^{'}=a\, \theta+ \, b\, \phi\; .$$
Because it is a linear combination of $\theta$ and $\phi$,
$\theta^{'}$ is a solution of Eq. (38). Taking the first, the
second and the third derivatives  of $S_0$ with respect to $x$
from Eq. (40), then replacing theme into Eq. (36), we find
\begin{eqnarray}
{\hbar^2 \over g_{11}}\, {\left[\left(\lambda\, {g_{11}\over {\sqrt{|g|}}}\right)^2-
W^2\right ]\over (\theta^{\, '2}+\phi^2)^2}\,
-{\theta^{'} \over (\theta^{\, '2}+\phi^2)}\,
\left[{\hbar^2 \over g_{11}}\,
{\partial^2 \theta^{'} \over \partial x^2}+\, {\hbar^2 \over \sqrt{|g|}}\, .
\right.
\hskip20mm\nonumber
\\
\left.
.\, {\partial \over \partial x}\left({\sqrt{|g|} \over g_{11}}\right)\,
{\partial \theta^{'} \over \partial x}+\,
{1 \over c^2}\, \left(m_o^2c^4-{E^2 \over g_{oo}}\right)
\right ]
-{\phi \over (\theta^{\, '2}+\phi^2)}\, .
\hskip18mm\nonumber
\\
.\, \left[{\hbar^2 \over g_{11}}\, {\partial^2 \phi \over \partial x^2}+
\, {\hbar^2 \over \sqrt{|g|}}\,
{\partial \over \partial x}\left({\sqrt{|g|} \over g_{11}}\right)\,
{\partial \phi \over \partial x}+
{1 \over c^2}\, \left(m_o^2c^4-{E^2 \over g_{oo}}\right)
\right ]=0\, . \hskip1mm
\end{eqnarray}
Taking account of Eq. (39) and since $\theta^{'}$ and $\phi$ are solutions of Eq. (38),
Eq. (41) is automatically satisfied. This demonstrate that
the expression (4) of the reduced action is the solution of the GQSHJE (Eq. (36)).

\vskip1.\baselineskip \noindent { {\bf 4.\ Gravitational Quantum
Law of Motion for \\ a stationary case and a static metric} }
\vskip0.25\baselineskip

Before studying the dynamical behaviour of a quantum particle
moving under the gravitational field, we introduce a quantum
gravitational transformation of the coordinate $x \to \hat{x}$
with which the GQSHJE giving in Eq. (36) takes the form of the
classical gravitational Hamilton-Jacobi equation (Eq. (37)) with
respect to the new coordinate $\hat{x}$. Such a transformation is
defined as
\begin{eqnarray}
\left({\partial x \over \partial \hat{x}}\right)^2=
\left\{1-{\hbar^2 \over 2}\left(\partial S_0 \over \partial x\right)^{-2}
\left[\{S_o,x\}-2\, \sqrt{g_{11} \over \sqrt{|g|}}\,
{\partial^2 \over \partial x^2}\left(\sqrt{\sqrt{|g|} \over g_{11}}\right)\, \right]\right\}
\nonumber\\
={g_{11} \over c^2}\, \left({\partial S_0 \over \partial x}\right)^{-2}\left(m^2_o\, c^4-\, {E^2 \over g_{oo}}\right)\; .
\hskip48mm
\end{eqnarray}
Here, we want to underline that the notion of a gravitational
quantum coordinate $\hat{x}$ is extrapolated from the notion of
the quantum coordinate already introduced by Faraggi and Matone
in Refs. \cite{FM1,FM2}.

Now, in order to derive the Gravitational quantum Law of motion, let us
introduce a Lagrangian of the form
\begin{equation}
L(x,\dot{x},a,b)=-m_0c^2 \sqrt{g_{oo}(x)+g_{11}(x)\, f(x,a,b)\, {\dot{x}^2 \over c^2}\;
}\; ,
\end{equation}
Where  $a$ and $b$ are two
constants of integration given in Eq. (4).
$f$ is a real function given by
\begin{equation}
f(x,a,b)=\left({\partial x \over \partial \hat{x}}\right)^{-2}\, =\,
{c^2 \over g_{11}}\, \left({\partial S_0 \over \partial x}\right)^{2}\left(m^2_o\, c^4-\, {E^2 \over g_{oo}}\right)^{-1}
\; .
\end{equation}
Using this expression of the Lagrangian, the least action principle leads to
\begin{eqnarray}
-{m_o\, c^2 \over 2}\, \left(1+{g_{11}\, f \over g_{oo}}{{\dot{x}}^2 \over c^2}\right)^{-{3 \over 2}}\,
\sqrt{g_{oo}}\, {d \over dt}\, \left({{\dot{x}}^2 \over c^2}{g_{11}\, f \over g_{oo}}\right)\, +
\hskip25mm\nonumber\\
m_o\, c^2\, \left(1+{g_{11}\, f \over g_{oo}}{{\dot{x}}^2 \over c^2}\right)^{-{1 \over 2}}\,
{d \over dt}\, (\sqrt{g_{oo}})=0\; ,
\end{eqnarray}
which, after integration, gives
\begin{equation}
E={m_o\, c^2\, g_{oo}(x)
\over
\sqrt{g_{oo}(x)+\, g_{11}(x)\, f(x,a,b)\, {{\dot{x}}^2\over c^2}}}\; \;  .
\end{equation}
$E$ is an integration constant representing the total energy of the quantum
particle moving in the presence of a gravitational field. Note that in the case
of a vanishing gravitational field ($g_{oo}=1$ and $g_{11}=-1$) and at the classical
limit ($\hbar \to 0, f(x,a,b) \to 1$), Eq. (46) reduces to the conservation equation
\begin{equation}
E={m_o\, c^2
\over
\sqrt{1-\, {{\dot{x}}^2\over c^2}}}
\end{equation}
well known in special relativity. Remark that taking the expression of \break
$f(x,a,b)$ from Eq. (44) in Eq. (46), we deduce the relation
\begin{equation}
\dot{x}\, .\, {\partial S_0 \over \partial x}=\, E-{m_o^2 c^4 \over E}\, g_{oo}(x)
\end{equation}
connecting between the momentum $\partial S_0 / \partial x$ and the velocity
of the particle. This relation represents the gravitational quantum law of motion.
Indeed, once the metric ($g_{oo}$ and $g_{11}$) of the 2D spacetime is determined, we can
plot the trajectories of the particle from Eq. (48)
after determining the solutions $\theta$ and
$\phi$ of the Klein-Gordon equation (Eq. (38)). For a vanishing gravitational field
Eq. (48) reduces to Eq. (10) (V(x) being vanish) representing the relativistic quantum
law of motion.

\vskip1.\baselineskip \noindent {{\bf 5.\ Hilbert-Einstein action
and the GQ Law of motion} } \vskip0.25\baselineskip

In the present section, we derive the GQ law of motion of a quantum particle moving
under the gravitational field in the general case (ie the field is not static and
the system is not stationary). For this aim, let us introduce the
Hilbert-Einstein action
\begin{equation}
S(x^{\mu})=-m_0c\int ds\, = -m_0c\int \sqrt{G_{\mu\nu}\dot{x}^{\mu}\dot{x}^{\nu}}\, ds\; ,
\end{equation}
where
\begin{equation}
ds^2=G_{\mu\nu}dx^{\mu}dx^{\nu}\; ,
\end{equation}
and $G_{\mu\nu}(x^{\alpha})$ are the components of the
gravitational quantum metric's tensor, and
$\dot{x}^{\mu}=dx^{\mu}/ds$. Writing expression (49) of the action
as follows
\begin{equation}
S(x^{\mu})= \int L(x^{\mu},\dot{x}^{\mu}) ds\; ,
\end{equation}
one can define the Lagrangian as
\begin{equation}
L(x^{\mu},\dot{x}^{\mu})=-m_0c
\sqrt{G_{\mu\nu}\dot{x}^{\mu}\dot{x}^{\nu}}\; .
\end{equation}
 \noindent Using expression (52) of the Lagrangian, the
least action principle leads to
\begin{equation}
{d^2 x^{\sigma} \over ds^2}+{\bf \Gamma_{\alpha\beta}^\sigma} {d x^{\alpha}
\over ds}{d x^{\beta} \over ds}=0\; , \hskip17mm \alpha, \, \beta=0,\,
1\; .
\end{equation}
where the ${\bf \Gamma_{\mu\nu}^\gamma}$ are the Christoffel
symbols given in Eq. (20), but where the metric tensor
$g_{\mu\nu}$ is replaced by $G_{\mu\nu}$. The action (49) is
invariant under reparametrization $s \to f (s)$. This gauge
symmetry leads to the constrained dynamics in the Hamiltonian
formulation \cite{Dirac}. The constraint reads
\begin{equation}
G^{\mu\nu}\, p_{\mu}\, p_{\nu}-m_0^2c^2=0\; ,
\end{equation}
where $p_{\mu}=\partial L/\partial \dot{x}^{\mu}$ are canonical momenta.

\noindent In the Hamilton-Jacobi theory the canonical momenta are
defined as
\begin{equation}
p_{\mu}={\partial S \over \partial x^{\mu}}\; .
\end{equation}
Replacing Eq. (55) into Eq. (54), we get
\begin{equation}
G^{\mu\nu}\, {\partial S \over \partial x^{\mu}}{\partial S \over
\partial x^{\nu}}-m_0^2c^2=0\; .
\end{equation}
Eq. (56) represents the Hamilton-Jacobi equation for a quantum
particle moving under the gravitational field. It must be
equivalent to Eq. (28). After comparison of Eqs. (28) and (56),
we deduce that the components of the metric tensor $G^{\mu\nu}$
must be written as
\begin{equation}
G^{\mu\nu}=g^{\mu\nu}\, f^{(\mu\nu)}
\end{equation}
Where $g^{\mu\nu}$ are the components of the metric tensor in the
missing of the quantum potential ($\hbar \to 0$), and

\begin{equation}
f^{(\mu\nu)}=\left[1-{\; \hbar^2 \over A}\, (\partial_{\mu}S\,
\partial_{\nu}S)^{-1}\, \left(\partial^{2}_{\mu\nu}\, A\, +
\, {1 \over g^{\mu\nu}\sqrt{|g|}}\, \partial_{\mu}\, (\sqrt{|g|}\,
g^{\mu\nu})\,
\partial_{\nu}\, A\, \right)\right]
\end{equation}
are the components of the metric tensor in the presence of a
quantum potential and the missing of the gravitational
interaction. It is useful to underline that the right hand side
of Eq. (57) do not support a summation over $\mu$ and $\nu$, it
is just a product of the components $g^{\mu\nu}$ and
$f^{(\mu\nu)}$. We note also that there is no summation in the
right hand side of Eq. (58).
 It is clear that when we take the classical limit
($\hbar \to 0$) the components $G^{\mu\nu}$ reduce to
$g^{\mu\nu}$.

Now, if one consider the stationary system and a static metric of
the space, he will find the results obtained in Sec. 4. So, the
expression (52) of the Lagrangian reduces to the expression (43),
and the dynamical equation (53) reduces to Eq. (45). In this case
the components $G^{\mu\nu}$ (Eq. (57)) reduce to
\begin{eqnarray}
G^{\mu\nu}=\begin{pmatrix}
      {{1\over g_{oo}} &0  \cr
        0&\left\{{1\over g_{11}}-{1\over g_{11}}{\hbar^2 \over 2}\left(\partial S_0 \over \partial
x\right)^{-2} \left[\{S_o,x\}-2\, \sqrt{g_{11} \over \sqrt{|g|}}\,
{\partial^2 \over \partial x^2}\left(\sqrt{\sqrt{|g|} \over
g_{11}}\right)\, \right]\right\} \cr}
 \end{pmatrix}\hskip-5mm\nonumber\\
 =\begin{pmatrix}
      {{1\over g_{oo}} &0  \cr
        0&{1\over g_{11}\, f(x,a,b)} \cr}
        \end{pmatrix}\; ,\hskip80mm
\end{eqnarray}
where $f(x,a,b)$ is given in Eq. (44). Replacing expression (59)
of the metric tensor into Eq. (52), the Lagrangian takes the form
given in Eq. (43). Then, using Eq. (55) of the canonical momenta
and taking into account $p_{\mu}=\partial L/\partial
\dot{x}^{\mu}$, we get
\begin{equation}
{\partial S \over \partial x^0}={1\over c}\, {\partial S \over
\partial t}=-{E\over c}={\partial L\over\partial
\dot{x}^{0}}=-{m_oc^2\, g_{oo}\, {dt\over ds} \over
\sqrt{g_{00}c^2\, {dt \over ds}^2+g_{11}f(x,a,b)\, \dot{x}^2 }}\;
,
\end{equation}
which can be written as
$$ E={m_oc^2\, g_{oo}\, \over
\sqrt{g_{00}+g_{11}f(x,a,b)\, \left({\dot{x}\over c}\right)^2 }}\;
.
$$
The last equation is equivalent to Eq. (46). It represents the
expression of the total energy of the particle for a stationary
system and a static metric.

The fact that the dynamical equations describing the motion of a
quantum particle under gravitation in the general case reduce to
the those describing the motion in the case of a stationary system
and a static metric is an important results. It indicates that the
generalization for a general system that we presented in this
section is correct, so the GQ law of motion given in this section
can be considered as the {\bf Universal Law of Motion}. Thus, in
2D, the geodesic equation (53) describes the trajectories of a
quantum particle moving under gravitation in a space with a
general metric and for a non-stationary systems. Such a system
has a Lagrangian of the form (52) and verify the GQHJE (Eq. (28)).

\vskip1.\baselineskip \noindent { {\bf 6.\ Conclusion} }
\vskip0.25\baselineskip

First, we would like to underline that in this article we do not
bring a description of the geometry of the 2D curved space under
gravitation. In other words, we do not specify the form of the
metric tensor $g_{\mu\nu}$ of the 2D timespace, we just exposed
the manner with which the 2D general relativity should be
connected with the 2D deterministic approach of quantum mechanics
presented in Refs.
\cite{BD1,BD2,BD3,Djama1,Djama2,Djama3,Djama4,Djama5}.

In the present paper, there is four main results. The first one
is the establishment of the GQHJE for general metric and
non-stationary case (Eqs. (28) and (29)). The second  result is
the establishment of the GQSHJE (Eq. (36)) for the stationary
case and a static metric. The third result is the elaboration of
the gravitational quantum law of motion for static metric and
stationary system  (Eqs. (43), (46) and (48)). The forth result is
the generalization of the GQ law of motion into general metric
and non-stationary system (Eqs. (52), (53), (54) and (56)). Note
that when we take the limit $\hbar \to 0$, all the established
equations reduce to those of the general relativity (GR), and
when we consider a vanishing gravitational potential, all the
established equations reduce to those of the Relativistic quantum
mechanics \cite{Djama1,Djama2}.

We would like to stress that the introduction of the gravitation
in our deterministic approach of quantum mechanics in
2-dimensions is an important and a daring step to connect the GR
and the deterministic quantum mechanics especially for the 4D
spacetime. We think that in order to generalize our deterministic
approach of quantum gravity into 4D, it is primordial to use the
Einstein-Hilbert action (Eqs. (49) and (51)). This suggestion
will be exposed in a preparing paper.

\vskip\baselineskip \noindent {\bf \ REFERENCES}
\vskip\baselineskip
%


\begin{enumerate}

\bibitem{BD1}
A. Bouda and T. Djama, {\it Phys. Lett.} A 285, 27 (2001);
quant-\break ph/0103071.

\bibitem{BD2}
A. Bouda and T. Djama, ; \textit{Physica scripta } 66 (2002)
97-104; quant-ph/0108022.

\bibitem{BD3}
A. Bouda and T. Djama, \textit{Phys. Lett.} A 296 (2002) 312-316;
quant-ph/0206149.

\bibitem{Djama1}
T. Djama, "Relativistic Quantum Newton's Law and photon
Trajectories"; quant-ph/0111121.

\bibitem{Djama2}
T. Djama, "Nodes in the Relativistic Quantum Trajectories and
Photon's Trajectories." ;quant-ph/0201003.

\bibitem{Djama3}
T. Djama, "The Relativistic Quantum Stationary Hamilton Jacobi
Equation for Particle with Spin 1/2.";
quant-ph/0311057.\\
T. Djama, "The Relativistic Quantum Law of motion for a Particle
with Spin $1/2$."; quant-ph/0311057.

\bibitem{Djama4}
T. Djama, "The 3D Quantum motion", quant-ph/0311082.

\bibitem{Djama5}
T. Djama, "3D Quantum Trajectories. Quantum orbits of the
Hydrogen's electron.", quant-ph/0404175.

\bibitem{FM1}
A.~E.~Faraggi and M.~Matone, {\it Int. J. Mod. Phys.} A 15, 1869
(2000); hep-th/9809127.

\bibitem{FM2}
A. E. Faraggi and M. Matone, {\it Phys. Lett.} A 249,180 (1998);
hep-ph/9801033.


\bibitem{Flo1}
E. R. Floyd, \textit{Found. Phys. Lett.} 9, 489 (1996).

\bibitem{Bohm1}
D. Bohm, Phys. Rev. 85, 166 (1952); 85, 180 (1952); D. Bohm and
J. P. Vigier, Phys. Rev. 96, 208 (1954).

\bibitem{Brog1}
L. de Broglie, Les incertitudes d'Heisenberg et
l'interpr{\'e}tation probabiliste de la m{\'e}canique ondulatoire
, (Gauthier-Villars, 1982), Chap. XII. L. de Broglie, Comp. rend.
183 , 447 (1926); 184 , 273 (1927); 185 , 380 (1927);

\bibitem{Flo2}
E. R. Floyd, Phys. Rev. D 34, 3246 (1986).


\bibitem{Mann1}
R.B. Mann, Found. Phys. Lett. 4 (1991) 425.

\bibitem{Mann2}
R.B. Mann, A. Shiekh, and L. Tarasov, Nucl. Phys. B341
(1990) 134.

\bibitem{Mann3}
J.D. Christensen and R.B. Mann, Class. Quant. Grav. (to be
published).

\bibitem{Mann4}
A.E. Sikkema and R.B. Mann, Class. Quantum Grav. 8 (1991)
219.

\bibitem{Mann5}
R.B. Mann and T.G. Steele, Class. Quant. Grav. 9 (1992) 475.

\bibitem{Mann6}
[17] R.B. Mann, S. Morsink, A.E. Sikkema and T.G. Steele, Phys.
Rev. D43 (1991) 3948.

\bibitem{Mann7}
[18] S.M. Morsink and R.B. Mann, Class. Quant. Grav. 8 (1991)
2257.

\bibitem{Mandal}
G. Mandal, A.M. Sengupta, and S.R. Wadia, Mod. Phys. Lett.
6 (1991) 1685.

\bibitem{Witt}
E. Witten, Phys. Rev. D44 (1991) 314.

\bibitem{Napp}
McGuigan M D, Nappi C R and Yost S A 1992 Nucl. Phys. B 375 421.

\bibitem{Call}
C. G. Callan, S. B. Giddings, J. A. Harvey, A. Strominger, Phys. Rev. D
45, (1992).

\bibitem{Mann8}
R.B. Mann and S.F. Ross Phys. Rev. D 47 (1993) 3312-3318.


\bibitem{Teo}
Perry M J and Teo E 1993 Phys. Rev. Lett. 70 2669.

\bibitem{Mann9}
Mann R B, Morris M S and Ross S F 1993 Class. Quantum Grav. 10 1477

\bibitem{Mann10}
Mann R B and Ross S F 1993 Class. Quantum Grav. 10 1405

\bibitem{EIN}
A. Einstein, Annalen Phys. 49 , 769 (1916).\\
Exact solutions of Einstein's field equations D. Kramer, H. Stephani, M. Maccullum
and E. Herlt (Cambridge University Press, 1980).

\bibitem{Dirac}
Dirac P A M 1964 Lectures on Quantum Mechanics (New York:Yeshiva
University Press ).



\bibitem{Dar}
S. K. Moayedi and F. Darabi, "Families of exact solutions of a 2D gravity model
minimally coupled to electrodynamics", arXiv: gr- \break qc/0012062.


\end{enumerate}
\end{document}